# Crystallographic dependent transport properties and oxygen issue in superconducting LiTi$_2$O$_4$ thin films

Y. L. Jia[1], G. He[1], H.S. Yu[1], W. Hu[1], Z. Z. Yang[1], J. A. Shi[1], Z. F. Lin[1], J. Yuan[1], B. Y. Zhu[1], K. Liu[2], L. Gu[1,3,4], H. Li[1,4], and K. Jin[1,3,4*]

[1]Beijing National Laboratory for Condensed Matter Physics, Institute of Physics, Chinese Academy of Sciences, Beijing 100190, China.
[2]Beijing Key Laboratory of Opto-electronic Functional Materials & Micro-nano Devices, Department of Physics, Renmin University of China, Beijing 100872, China.
[3]Collaborative Innovation Center of Quantum Matter, Beijing, 100190, China
[4]School of Physical Sciences, University of Chinese Academy of Sciences, Beijing 100190, China

Abstract: A systematic study is performed on the spinel oxide, *i.e.* LiTi$_2$O$_4$ thin films oriented along [111]-, [110]-, and [001]-directions, to reveal the crystallographic dependence of transport properties. With decreasing temperature, the LiTi$_2$O$_4$ displays roughly identical onset temperatures of twofold symmetry of in-plane angular dependent magnetoresistivity (AMR) (at ~ 100 K), crossover from negative- to positive- magnetoresistance (at ~ 50 K), and coherence length in the superconducting state. While, the twofold symmetry in AMR itself suggests anisotropic electron scattering. The superconducting transition temperature ($T_c$) seems insensitive to the lattice parameter. Moreover, the spherical aberration-corrected scanning transmission electron microscopy (Cs-STEM) discloses that oxygen vacancies exist in the LiTi$_2$O$_4$ films. These oxygen vacancies cause the change of lattice but show little influence on superconductivity, differing from high-$T_c$ cuprates where subtle variation of oxygen way lead to a significant change in superconductivity.

The metallic spinel lithium titanate $LiTi_2O_4$ is the only known spinel oxide superconductor, which was discovered by Johnston et al. in 1973 [1]. The critical transition temperature ($T_c$) of superconducting $LiTi_2O_4$ is as high as 13.7 K [2]. Previous work on polycrystalline samples has disclosed that $LiTi_2O_4$ is a BCS s-wave superconductor with mediate electron-phonon coupling [3,4]. Meanwhile, owing to the mixed-valence of Ti ions in frustrate Ti sublattice, $LiTi_2O_4$ exhibits complex electron-electron correlations via spin fluctuations, supported by the resonant inelastic soft-x-ray scattering [5], nuclear magnetic resonance [6] and magnetic susceptibility measurements [7]. Moreover, the substitution of Li for Ti will cause a superconductor-to-insulator transition, *e.g.* from superconducting $LiTi_2O_4$ to insulating $Li_4Ti_5O_{12}$ [8-10]. The relationship between the content of Li ions, O vacancies and the superconductivity in $LiTi_2O_4$ is yet to be clarified, and it is necessary to perform comprehensive investigation on single crystalline samples.

Great efforts have been made on improving the quality of $LiTi_2O_4$ material in the past four decades [7,11,12], but it is failed to get stabilized $LiTi_2O_4$ single crystals. On one hand, $Ti^{3+}$ ions are unstable in the oxygenation circumstance [12]. On the other hand, $Li^+$ ions are easy to diffuse from 8a to 16c sites [2]. Alternatively, high quality $LiTi_2O_4$ thin films may overcome these drawbacks and be a good candidate. Recently, electrical transport and tunneling spectra measurements on [001]-oriented high quality single crystal $LiTi_2O_4$ thin films have revealed that an orbital related state appears below ~ 50 K and strong spin fluctuations exist in the temperature range of 50 ~ 100 K [13]. Unlike the unconventional high-$T_c$ cuprates and Fe-based superconductors in quasi two-dimensional structure [14,15], $LiTi_2O_4$ belongs to a space group of $F$d3m, a cubic structure. Therefore, it is interesting to know whether electronic structure of $LiTi_2O_4$ behaves isotropic or anisotropic. In fact, tunneling spectra have disclosed anisotropy of electron-phonon coupling for $LiTi_2O_4$ films of different orientations [16], whereas a detailed transport study is still lacking.

In this paper, we have grown the high quality $LiTi_2O_4$ thin films on [001]-, [110]- and [111]-oriented substrates. All $LiTi_2O_4$ films show almost the same $T_c$ (~ 11 K) with transition width of less than 0.5 K. For these samples of different orientations, a crossover from positive to negative magnetoresistance (MR) happens at ~ 50 K and a twofold symmetry in in-plane angular dependent magnetoresistivity (AMR) shows up below ~ 100 K, indicating roughly isotropic band structure but anisotropic scattering in this system. Furthermore, unexpected oxygen vacancies have been detected by the spherical aberration-corrected scanning transmission electron microscopy techniques (Cs-STEM), which lead to reduction of $Ti^{4+}$ ions. Interestingly, we find that the $T_c$ seems insensitive to the variation of lattice constant as long as the samples are in the metallic regime.

The $LiTi_2O_4$ thin films are deposited with different oxygen pressures and substrates by pulsed laser deposition (PLD) technique with thickness of ~ 200 nm in the present work. Before the deposition, the $MgAl_2O_4$ (MAO) substrates are annealed at 1000 °C for 5 hours in air [8], to obtain smooth surface. The sintered $Li_4Ti_5O_{12}$ ceramic target is used to grow both $LiTi_2O_4$ and $Li_4Ti_5O_{12}$ films with a KrF excimer laser (λ = 248 nm), pulse frequency of 4 Hz, energy density of 1.5 J/cm², and deposition temperature of ~ 700 °C. Pure $LiTi_2O_4$ can be obtained in vacuum better than $10^{-6}$ Torr, whereas $Li_4Ti_5O_{12}$ forms in oxygen pressure of higher than $10^{-4}$ Torr [17]. After the deposition, both $LiTi_2O_4$ and $Li_4Ti_5O_{12}$ thin films are quenched to room temperature in vacuum. The structure is indicated by X-ray diffraction (XRD). The electrical resistivity is measured by four-probe method.

The $\theta$-$2\theta$ X-ray diffraction (XRD) spectra of [001]-oriented samples grown under different oxygen pressures ($P_{O2}$) are shown in Fig. 1(a). It is clear that with increasing $P_{O2}$, the (008) peak of films gradually moves from ~ 94° to ~ 95.3°, corresponding to a transformation from $LiTi_2O_4$ to $Li_4Ti_5O_{12}$ phase. The c-axis lattice constant versus $P_{O2}$ is shown in Fig. 1(b). At $P_{O2}$ < $10^{-4}$ Torr, it decreases linearly with increasing oxygen pressure (Fig. 1(b)) and $T_c$ remains almost constant (Fig. 1(c)). However, when $P_{O2}$ is higher than $10^{-4}$ Torr, the lattice constant stays unchanged, and the superconducting $LiTi_2O_4$ with $T_c$ ~ 11 K suddenly goes to an insulating $Li_4Ti_5O_{12}$ phase. In Fig. 2(a), we present XRD spectra of $LiTi_2O_4$ films on different substrates, *i.e.* [001]-, [110]- and [111]-oriented $MgAl_2O_4$ (MAO), [001]- and [111]-oriented $SrTiO_3$ (STO), as well as [001]-oriented $KTaO_3$ (KTO) single crystal substrates. Structural analyses indicate that these films are in pure phase. In Fig. 2(b), $\varphi$-scan measurement in (404) plane displays a fourfold symmetry with uniformly distributed peaks. Fig. 2(c) shows the reciprocal space mapping of $LiTi_2O_4$ film, which indicates an epitaxial growth as well.

Fig. 3(a) shows the $\rho$-$T$ curves of LiTi$_2$O$_4$ thin films on different substrates. Here our samples still keep sharp superconducting transition after several months in air. Although the values of $T_c$ in LiTi$_2$O$_4$ films on different substrates are almost the same, their residual resistivity ratios (RRR) are quite different, suggesting that the superconductivity in LiTi$_2$O$_4$ is not sensitive to disorders, *e.g.* oxygen vacancies [13]. Fig. 3(b) shows the normalized d$\rho_{xx}$/d$T$ of LiTi$_2$O$_4$ films along different orientations. All these curves can be fitted with the Fermi liquid theory, *i.e.* d$\rho$/d$T$ ~ $T$ (red solid line) below 100 K. The field dependence of resistance on [001]- and [110]-oriented LiTi$_2$O$_4$/MAO films substrates are shown in Fig. 4(a) and 4(b). We define the upper critical field ($H_{c2}$) at 90% of the normal state resistivity ($R_n$) near $T_c$ and plot it against the temperature. As shown in Fig. 4(c), the $H_{c2}(T)$ curves do not show clear anisotropy. That is, the Ginzburg-Landau coherence length $\xi_{GL}(T)$ is basically identical, *i.e.* ~ 4.5 nm (Fig. 4(d)), calculated from $\xi_{GL} = \left(\frac{\phi_0}{2\pi H_{c2}}\right)^{1/2}$ with $\phi_0$ the fluxoid quantum.

Fig. 5(a) presents the AMR measurements at 9 T field. The film was rotated around the out-of-plane axis with $H$ parallel to the film surface. We define AMR = $\rho(\theta)$-$\rho_{min}$ and plot it as a function of $\theta$ for the films with different orientations. An obvious twofold symmetry of AMR can be observed for [110]-, [111]- as well as [001]-oriented LiTi$_2$O$_4$ samples. Fig. 5(b) shows the amplitude of anisotropic AMR, *i.e.* $\rho_{max}$ - $\rho_{min}$, at different temperatures, which shows a kink at 50 ± 10 K and disappears above ~ 100 K. We also draw the field dependence of the normal state resistivity at 30 K (solid symbols) and 70 K (open symbols) for [001]-, [110]- and [111]-oriented LiTi$_2$O$_4$ samples (see Fig. 5(c)). The MR of different oriented LiTi$_2$O$_4$ samples is positive at 30 K, but negative at 70 K. The temperature dependences of MR at 7 T for different oriented LiTi$_2$O$_4$ films are shown in Fig. 5(d). A crossover from positive to negative MR at ~ 50 K is observed. The onset temperature of anisotropic AMR and the sign change in MR are consistent with previous report on [001]-oriented samples [13].

We first summarize our major results, that is, for LiTi$_2$O$_4$ films of different orientations electrical transport properties are crystalline-independent, *e.g.* consistent onset temperatures of two-fold symmetry in AMR and crossover of magnetoresistance from negative to positive in the normal state, as well as roughly identical coherence length in the superconducting state. However, the superconducting transition temperature seems insensitive to the lattice constant in metallic $LiTi_2O_4$ films.

Previous work on [001]-oriented $LiTi_2O_4$ has disclosed that the positive MR and two-fold symmetry of AMR plausibly originate from an orbital-related state below 50 K, and the negative MR stems from spin-orbit fluctuations, which should also work for the [110]- and [111]-oriented films. That is, the electron scattering could be anisotropic in magnetic field and vary with temperature. While, the trajectories of electron motion in fields seem almost crystalline-independent, where the effective mass is isotropic and agreement with the band calculations [18,19]. This is not surprise because the $LiTi_2O_4$ has a 3D structure. In addition, we have observed anisotropic tunneling spectra in zero field, where phonon modes are clearly observed in the [110]- and [111]-oriented films but not the [001]-oriented ones [16]. Compared to the high-$T_c$ cuprates, quasi 2D materials are highly anisotropic in both the band structure and electronic correlations, our results reveal that the $LiTi_2O_4$ has roughly isotropic band structure but anisotropic electron scattering and electron-phonon coupling.

Furthermore, phonon modes have been detected in high $T_c$ cuprates via ARPES [20] and STM [21], but they seem not to contribute to the electrical resistivity, which always shows either a Fermi liquid ($\rho \sim T^2$) or strange metal behavior ($\rho \sim T$), *etc* [22]. In $LiTi_2O_4$, electron-phonon coupling is strong but the electrical resistivity still behaves as a Fermi liquid (Fig. 3(b)). Why the electron-electron scattering, rather than phonon scattering, dominates the transport requiring further investigations. Nevertheless, $LiTi_2O_4$ is a good candidate to address this open issue and may bridge the 3D conventional superconductors and the 2D novel superconductors. Another intriguing issue in present work is that the $T_c$ seems insensitive to the lattice constant, in which we know for high-$T_c$ cuprates and Fe-based superconductors, tine change in lattice constant always causes influence on $T_c$ [23,24].

In order to clarify this issue, we performed the Cs- STEM on our $LiTi_2O_4$ films as shown in Fig. 6. The oxygen vacancies exist in $LiTi_2O_4$ films as cluster. We select two different regions (blue and red frame) in the same film (Fig. 6(a)).

From the line profiles corresponding to the rectangle area 1-4 in the enlarged and filtered ABF image in Fig. 6(b)-(c), oxygen vacancies are distributed at two occupations, marked as O1 and O2 in the sketch of lattice, respectively. Interestingly, these oxygen vacancies are ordered along certain direction as shown in Fig. 6(d)-(g). The existence of oxygen vacancies lead to more electrons in LiTi$_2$O$_4$ samples, which induce the reduction of Ti$^{4+}$ ions, i.e. Ti$^{4+}$ to Ti$^{3+}$, as confirmed by the high-angle annular-dark-field (HAADF) image and the electron energy-loss spectroscopy (EELS) of LiTi$_2$O$_4$/MAO in Fig. 7. The asterisks marked EELS spectra of Ti $L_{2,3}$ show a slight variation in $e_g$-$t_{2g}$ band splitting, implying there is a higher Ti$^{4+}$/Ti$^{3+}$ ratio in corresponding areas in the LiTi$_2$O$_4$ film. Meanwhile less oxygen vacancies exist in the same areas, which is revealed by EELS spectra of O K edge. Both the STEM image and EELS spectra manifest that oxygen vacancies distribute inhomogeneously in our LiTi$_2$O$_4$ film.

With substituting of Li for Ti, Li$_{1+x}$Ti$_{2-x}$O$_4$ holds the onset transition temperature of superconductivity for $x < 1/3$, and seems suddenly to be an insulator at $x = 1/3$ (*i.e.* Li$_4$Ti$_5$O$_{12}$) [2]. The uncontinuous change in onset $T_c$ means phase separation between superconducting LiTi$_2$O$_4$ ($x = 0$) and insulating Li$_4$Ti$_5$O$_{12}$. In other words, with decreasing Li superconducting LiTi$_2$O$_4$ islands emerge in insulating Li$_4$Ti$_5$O$_{12}$ background, and a broad transition is thus observed in mixture of these two phases [7]. However, our superconducting films deposited in different oxygen pressures always show sharp transition, *i.e.* no two phase coexistence.

Compared to Li$_4$Ti$_5$O$_{12}$ in which the oxygen sites are fully occupied [25,26], remarkable oxygen vacancies exist in the superconducting LiTi$_2$O$_4$ films. In Fig. 1(c), the residual resistivity of LiTi$_2$O$_4$ turns better in higher vacuum, which means fewer disorders (*e.g.* oxygen vacancies) in sample with larger *c*-axis. Therefore, there are more and more disorders with increasing oxygen pressure during the deposition, but the superconducting transition remains sharp until the transition to Li$_4$Ti$_5$O$_{12}$. Hence, our work suggests that the superconductivity in LiTi$_2$O$_4$ is insensitive to oxygen vacancies. The increase of the number of carriers and lattice constant will reduce the bandwidth and increase the density of states near the Fermi level. Therefore, it is expected to enhance $T_c$ [27]. Then more questions arise: why there are fewer disorders (oxygen vacancies) in higher vacuum and why doped electrons by these oxygen vacancies do not affect $T_c$? These opening questions deserve further investigations beyond this work.

In conclusion, high quality $LiTi_2O_4$ films were successfully synthesized on various substrates and along different orientations. A careful study of electrical transport properties on [001]-, [110]-, and [111]-oriented samples discloses identical onset temperatures of two-fold symmetry in AMR (~ 100 K) and crossover from negative to positive MR (~ 50 K) of different orientations. While, the anisotropic AMR in $LiTi_2O_4$ films indicates an anisotropic electron scattering. Meanwhile, unexpected oxygen vacancies are found to exist in the superconducting $LiTi_2O_4$ based on our Cs-STEM results, which result in the reduction of Ti ions. We also find that the $T_c$ seems to be insensitive to out-of-plane lattice constant, implying that these oxygen vacancies are not the key role in superconductivity. In order to approach the nature of $LiTi_2O_4$, further investigations are required to solve intriguing issues such as that more oxygen vacancies come out under lower vacuum deposition and do not influence on $T_c$.

*Electronic address: kuijin@iphy.ac.cn

**Acknowledgements:** We thank Z.M. Fu for crucial discussion and L.H. Yang for technique support. This work was supported by the National Key Basic Research Program of China (Grant no. 2015CB921000), the National Natural Science Foundation of China (Grant no. 11474338 ), the Open Research Foundation of Wuhan National High Magnetic Field Center (Grant no. PHMFF2015008 ), and the Strategic Priority Research Program (B) of the Chinese Academy of Sciences (Grant no. XDB07020100). K.L. was supported by the Fundamental Research Funds for the Central Universities, and the Research Funds of Renmin University of China (14XNLQ03). L.G. was supported by National Program on Key Basic Research Project (2014CB921002) and The Strategic Priority Research Program of Chinese Academy of Sciences (Grant No. XDB07030200) and National Natural Science Foundation of China (51522212, 51421002, 51332001).

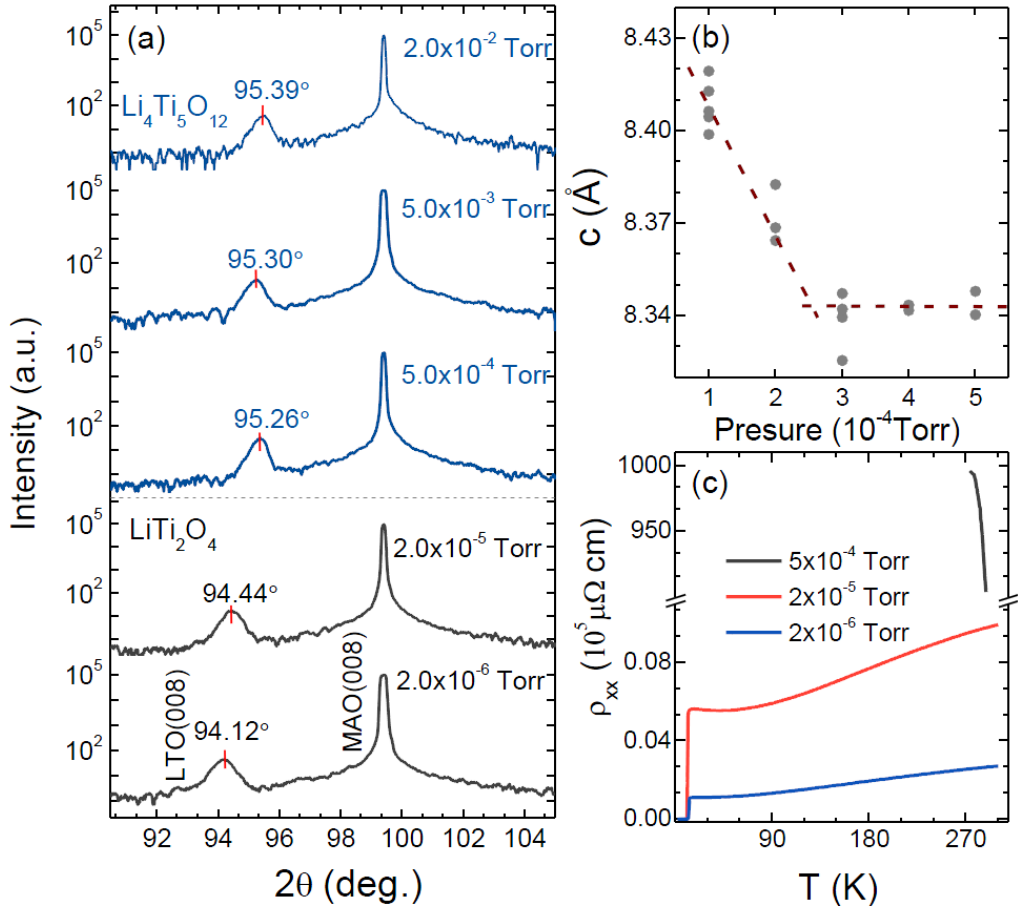

FIG. 1. (Color online) (a) The $\theta$-$2\theta$ X-ray diffraction (XRD) spectra of $LiTi_2O_4$ thin films grown at different $P_{O2}$. At $P_{O2} < 10^{-4}$ Torr, the $LiTi_2O_4$ phase is formed. The $Li_4Ti_5O_{12}$ phase is obtained when the $P_{O2}$ is up to $10^{-4}$ Torr. (b) The lattice constant of $c$-axis vs. $P_{O2}$. (c) $\rho$-$T$ curves of $LiTi_2O_4$ thin films with different $P_{O2}$. As $P_{O2}$ increases, the normal state resistivity for the films becomes larger, and the film turns to insulator when $P_{O2}$ reaches $5 \times 10^{-4}$ Torr.

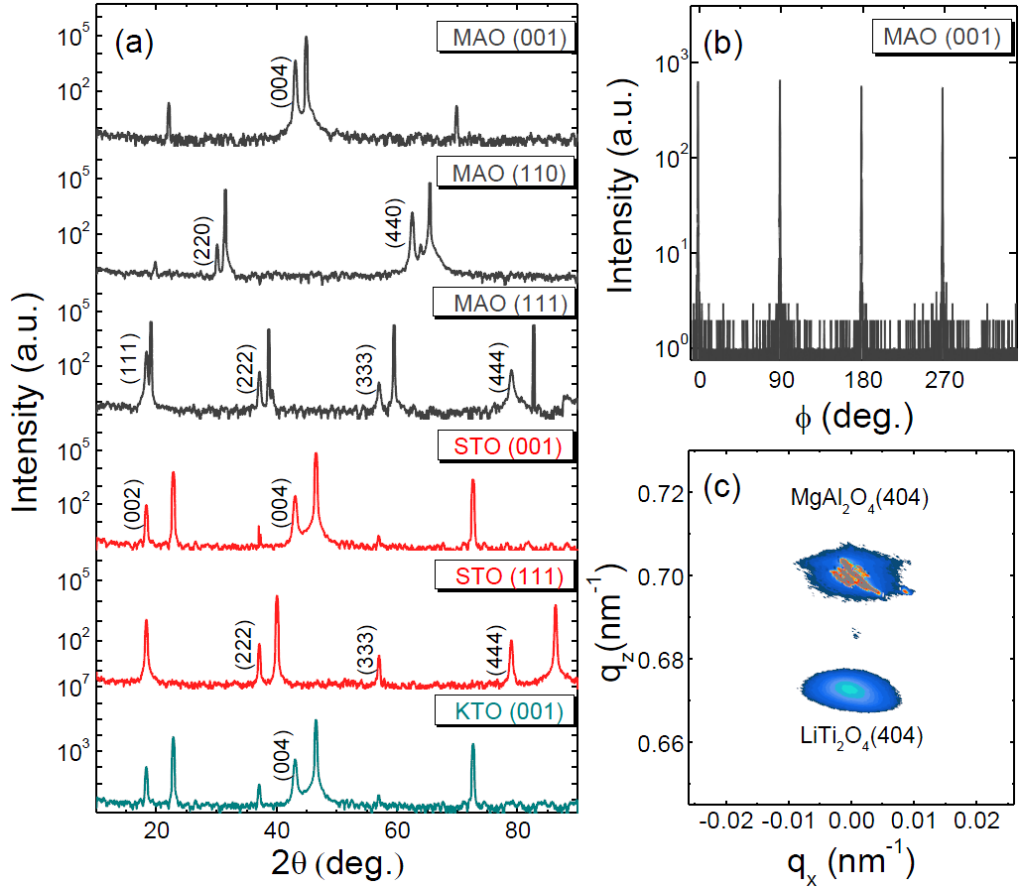

FIG. 2. (Color online) (a) The $\theta$-$2\theta$ XRD spectra of epitaxial LiTi$_2$O$_4$ films grown on different substrates. (b) The $\varphi$-scan measurement of LiTi$_2$O$_4$/MgAl$_2$O$_4$ [001] in the [404] reflection. Four peaks are uniformly distributed. (c) The reciprocal space mapping of LiTi$_2$O$_4$ [404] and MgAl$_2$O$_4$ [404] peaks.

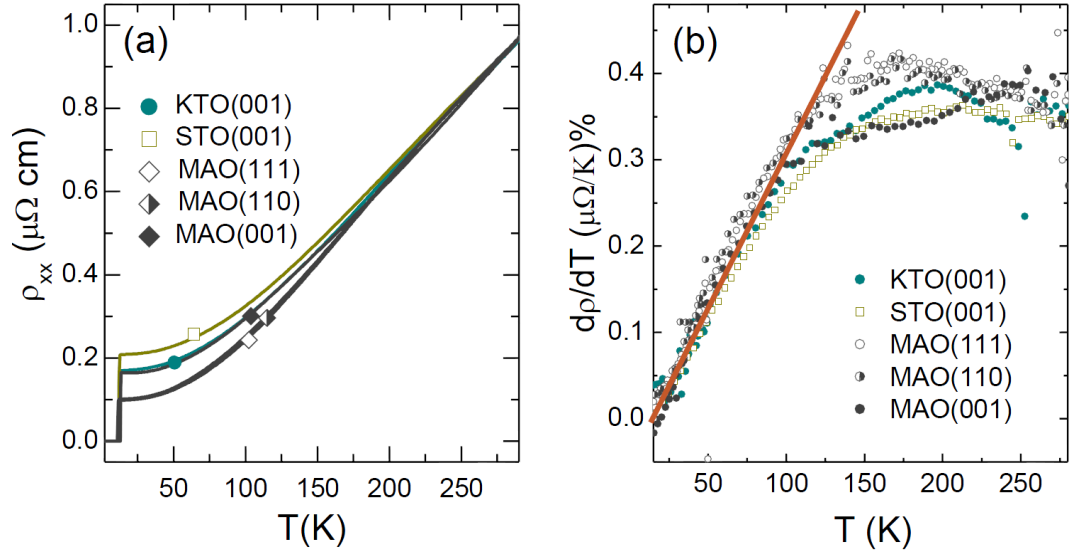

FIG. 3. (Color online) (a) The ρ-T curves of LiTi$_2$O$_4$ thin films grown on different substrates. Circle and square symbols represent films on KTaO$_3$ and SrTiO$_3$ substrate, respectively. Rhombus symbols represent films on MgAl$_2$O$_4$ substrates of different orientations. The values of $T_c$ are nearly the same. (b) The temperature dependence of the derivative of resistivity in [001], [110] and [111] LiTi$_2$O$_4$ films.

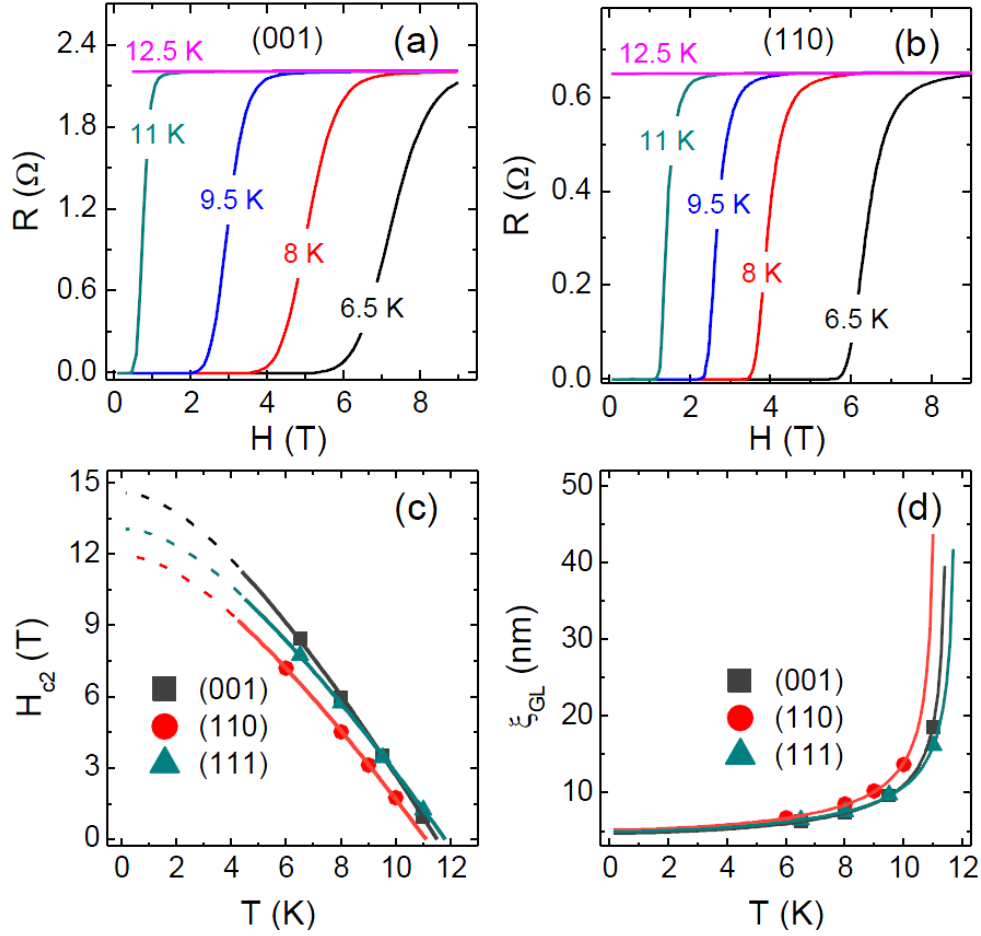

FIG. 4. (Color online) (a) (b) The field dependence of the in-plane (B⊥ab plane) magnetoresistivity of the [001] and [110] oriented samples around $T_c$. (c) Upper critical fields defined as the values of 90% of normal state resistivity ($R_n$) near $T_c$ plotted against the temperature. The experimental data (solid dots) can be fitted by $H_{c2}(T)= H_{c2}(0)[1- (T/T_c)^2]/[1+(T/T_c)^2]$ derived from the Ginzburg-Landau theory. (d) The temperature dependence of the coherence length $\xi_{GL}(T)$ for different oriented samples.

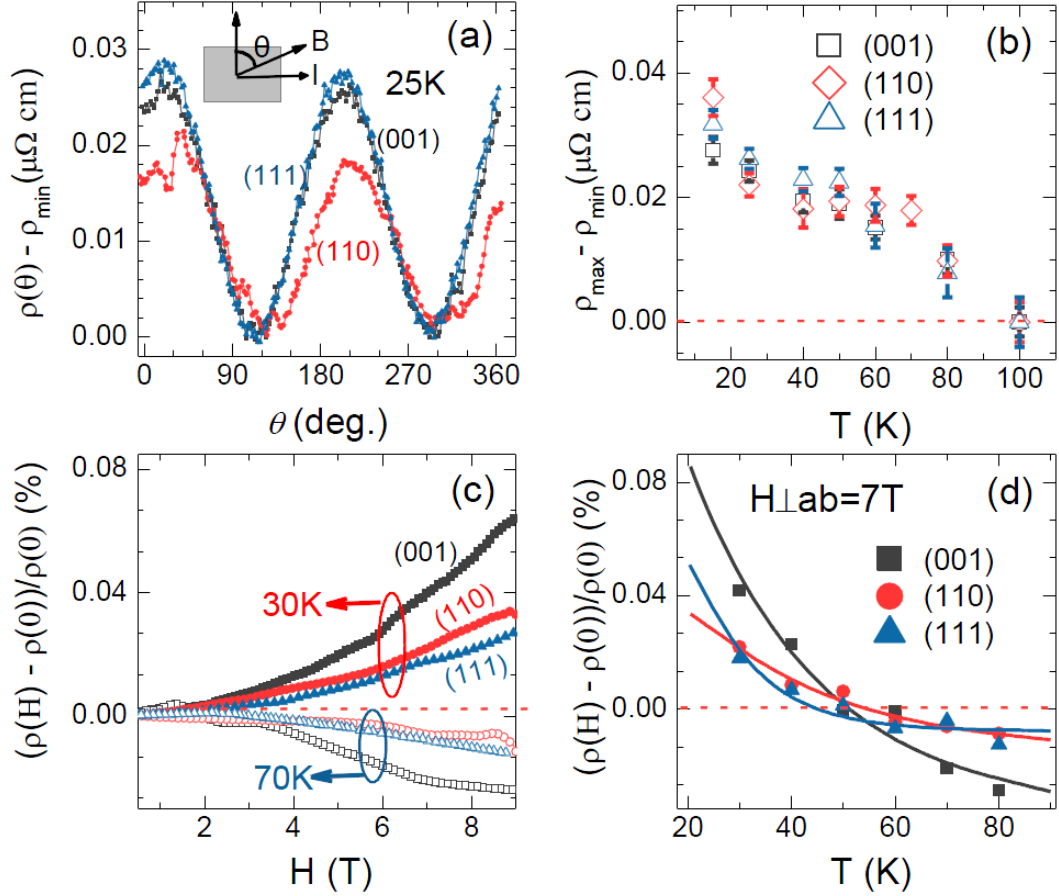

FIG. 5. (Color online) (a) The in-plane angular MR of [001]-, [110]- and [111]- oriented LiTi$_2$O$_4$ thin films at 25 K. (b) The temperature dependence of the amplitudes of the oscillation in AMR ($\rho_{max}$-$\rho_{min}$) for different oriented LiTi$_2$O$_4$ thin films. With decreasing temperature, the LiTi$_2$O$_4$ displays roughly identical onset temperatures of twofold symmetry in AMR at 100 K. (c) The field dependence of the in-plane MR at 30 K (solid symbols) and 70 K (open symbols) of different oriented LiTi$_2$O$_4$ thin films. (d) The fitted lines for MR at different temperature (solid symbols). With decreasing temperature, the signs of MR change from negative to positive around ~50 ± 10 K.

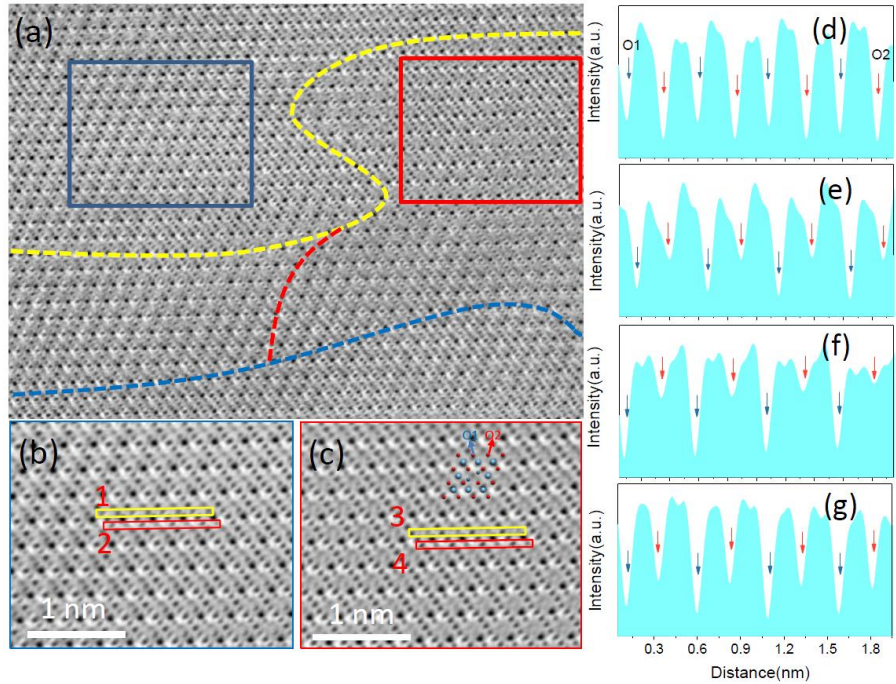

FIG. 6. (Color online) (a) The annular-bright-field (ABF) image of LiTi$_2$O$_4$ thin film. The interfaces of these phases are marked by dashed lines. (b) (c) The enlarged and filtered ABF image of the area marked with blue and red frame in (a), respectively. (d)-(g) The line profiles corresponding to the rectangle area 1-4 in (b) and (c). The O1 (blue arrow) and O2 (red arrow) represent the location of oxygen of LiTi$_2$O$_4$ structure diagram in (c), respectively.

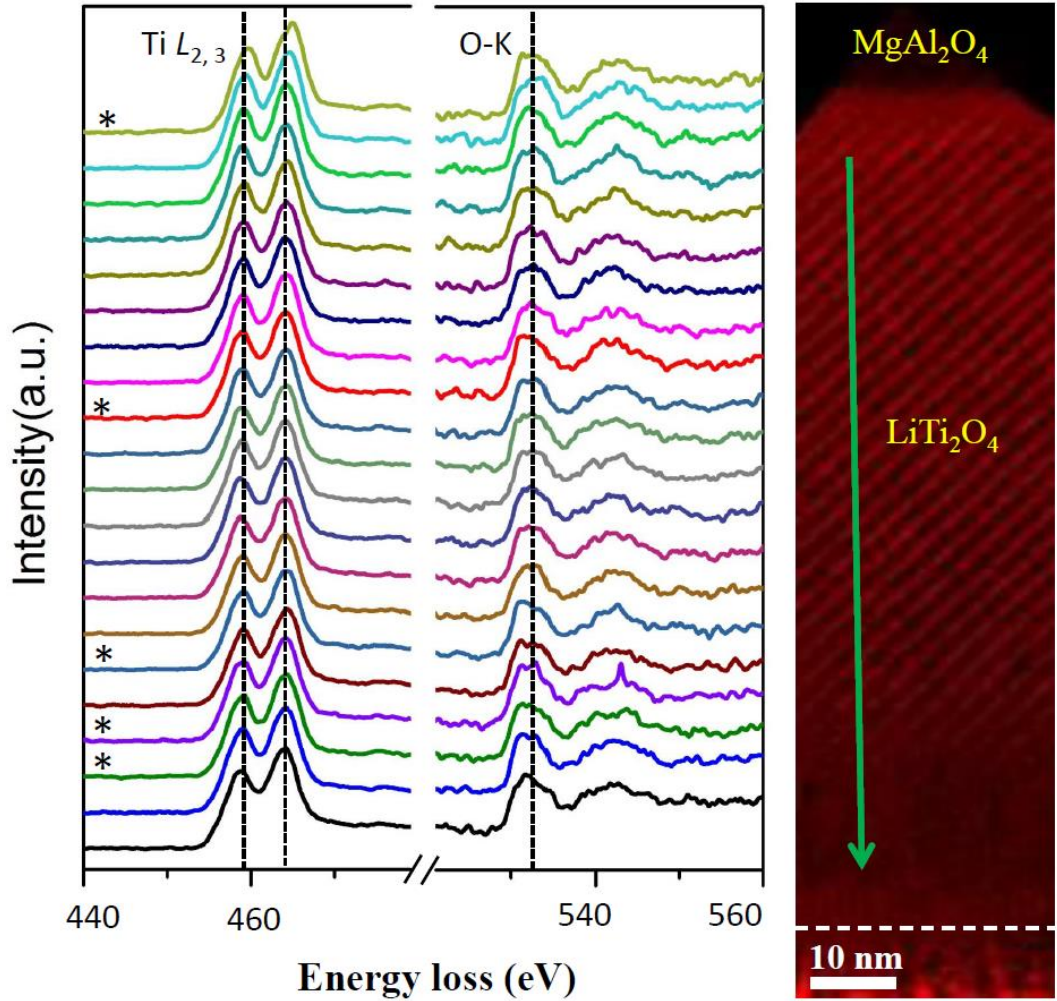

FIG. 7. (Color online) The electron energy-loss line-scan spectra (EELS) (left) and the high-angle annular-dark-field (HAADF) image (right) of LiTi$_2$O$_4$/MAO sample. Ti $L_{2,3}$ and O K edges are acquired simultaneously along the marked arrow in the HAADF image. A slight variation indication of $e_g$-$t_{2g}$ band splitting can be seen in the asterisks marked EELS spectra of Ti $L_{2,3}$, which indicates a high Ti$^{4+}$/Ti$^{3+}$ ratio in corresponding areas in the LiTi$_2$O$_4$ film. In addition, the marked EELS spectra also show different shape of O K edge due to less oxygen vacancies in these areas.